\documentclass[aps,prb,showpacs,twocolumn]{revtex4}
\usepackage{graphicx}
\usepackage{bm}
\begin{document}
\title{Nonlinear theory of fractional microwave-induced magnetoresistance oscillations\\ 
in a dc-driven two-dimensional electron system}
\author{X. L. Lei}
\affiliation{Department of Physics, Shanghai Jiaotong University,
1954 Huashan Road, Shanghai 200030, China}

\begin{abstract}

Microwave-induced nonlinear magnetoresistance in a dc-driven two-dimensional electron system 
is examined using a multi-photon-assisted transport scheme 
direct controlled by the current. 
It is shown that near the 2nd subharmonic of the cyclotron resonance, 
the frequency of the resistivity oscillation with the magnetic-field-normalized current-density 
is double that at the cyclotron resonance and its harmonics,
in excellent agreement with recent experimental findings by Hatke {\it et al.}
[Phys. Rev. Lett. {\bf 101}, 246811 (2008)]. 
The current-induced alternative emergence of resonant two-photon and single-photon processes
is responsible for this frequency doubling.
Near the third subharmonic of the cyclotron resonance, 
the current-induced consecutive appearance of resonant 0-/3-photon, two-photon, 
and single-photon processes
may lead to the frequency tripling of the resistivity oscillation.

\end{abstract}


\maketitle

\section{Introduction}

A number of extraordinary  magnetotransport phenomena of irradiated two-dimensional (2D) electrons 
in very high Landau levels (LLs), especially the microwave-induced resistance oscillation (MIRO) 
and the zero-resistance state, has  been the focus of 
intensive
experimental \cite{Zud01,Ye,Mani02,Zud03,Dor03,Yang03,Zud04,Mani04,Willett,Mani05,
Dor05,Stud,Smet05,Yang06} 
and theoretical \cite{Ryz-1970,Ryz03,Shi,Durst,Lei03,Vav04,Dmitriev03,DGHO05,Torres05,
Ng05,Ina-prl05,Kashuba,Andreev03,Alicea05,Auerbach05,Mikhailov04} studies in the past few years.

Under the irradiation of a moderate microwave of frequency $\omega$, the most striking resistance
oscillations show up in the linear photoresistivity as a function of the inverse magnetic field $1/B$, 
featuring the periodical appearance of peak-valley pairs around $\omega/\omega_c=1,2,3...$  
($\omega_c=eB/m$ is the cyclotron frequency and $m$ the electron effective mass). 
The cyclotron resonance and its harmonics are node points of peak-valley pairs,
where the photoresistivity vanishes. 
In addition, secondary peak-valley pairs 
were also observed in the linear resistance near certain fractional values,
$\omega/\omega_c=1/2, 1/3, 2/3$, and 3/2. These fractional microwave-induced resistance oscillations
were referred to real multiphoton participating processes.\cite{Lei06} 
They were also referred to higher harmonic effect,\cite{Dor03}
the resonant series of consecutive single-photon transition,\cite{Dor07,Pech07} 
or microwave-induced sidebands.\cite{Dmitriev07}
Despite different scenarios have been proposed for these fractional MIROs 
of linear photoresistivity, they have so far been theoretically demonstrated to clearly appear 
in a real system only within the multiphoton-assisted scattering model.\cite{Lei06} 

Another notable effect recently discovered is the significant role of a dc current.
A relatively weak current alone, without microwave radiation, can also induce substantial magnetoresistance 
oscillations and zero-differential resistance states.\cite{Yang02,Bykov05,WZhang07,JZhang07,
Bykov07,Lei07-1,Vav07,WZhang08,Lei08-1}
Simultaneous application of a finite dc current and a microwave radiation leads to very  
interesting and complicated oscillatory behavior of resistance and 
differential resistance.\cite{WZhang07-2,Lei07-2,Auerbach07,Hatke08,Lei08-2} 
These facts clearly indicate that
 microwave-irradiated nonlinear magnetotransport is remarkably different from the linear one.  
So far experimental and theoretical investigations on nonlinear MIROs have been carried out
in the magnetic field range lower than the cyclotron resonance and never anticipated any
anomaly elsewhere. Very recent experiments by Hatke {\it et al.} \cite{Hatke08-2} disclosed sharply distinct 
nonlinear transport behavior in the subharmonic fields of the cyclotron resonance.
Particularly, they found that at the second cyclotron resonance subharmonic, $\omega/\omega_c=1/2$,
the differential resistivity oscillates as a function of the current density in the same way 
as that at the cyclotron resonance, $\omega/\omega_c=1/2$, despite a factor of two difference in magnetic fields.
In other words, at $\omega/\omega_c=1/2$ the frequency of the resistivity oscillation with the 
magnetic-field-normalized current-density 
is twice that at the cyclotron resonance and its harmonics.
This is in vast contract with the observed behavior of the resistivity oscillations
in the magnetic field regimes of cyclotron resonance and its harmonics ($\omega/\omega_c\geq 1$). 
A nonlinear theoretical analysis capable of identifying the specific mechanism responsible for fractional MIROs 
and frequency doubling is highly desirable.

This paper presents such an  
analysis on microwave-induced nonlinear magnetotransport in high Landau levels in GaAs-based 2D heterosystems.
With the microscopic balance-equation scheme direct controlled by the current,
we are able to constitute a unified description for both the cyclotron resonance harmonic and subharmonic regimes 
of 
the magnetic fields and predict the resistivity oscillations with applied current density 
in excellent agreement with the experimental observation in both regimes. 

\section{Formulation}

Our examination is based on a current-controlled scheme of photon-assisted transport,\cite{Lei03}
which deals with a 2D system of short thermalization time having $N_{s}$ electrons in a unit area of 
the $x$-$y$ plane. These electrons, subjected to a uniform magnetic field ${\bm B}=(0,0,B)$ 
in the $z$ direction and scattered by random impurities and by phonons in the lattice, 
perform an integrative drift motion when an incident microwave electric field ${\bm E}_{\rm is}\sin \omega t$
irradiates on and a current flows in the plane.
In terms of the center-of-mass (c.m.) momentum and coordinate defined as 
${\bm P}\equiv\sum_j {\bm p}_{j\|}$ 
and ${\bm R}\equiv N_{s}^{-1}\sum_j {\bm r}_{j}$  
with ${\bm p}_{j\|}\equiv(p_{jx},p_{jy})$ and ${\bm r}_{j}\equiv (x_j,y_j)$
being the momentum and coordinate of the $j$th electron in the 2D plane,
and the relative electron momentum and coordinate 
${\bm p}_{j\|}'\equiv{\bm p}_{j\|}-{\bm P}/N_{s}$ and 
${\bm r}_{j}'\equiv{\bm r}_{j}-{\bm R}$, 
the Hamiltonian of this system can be written 
as the sum of a c.m. part $H_{\rm c.m.}$ and
a relative electron part $H_{\rm er}$,\cite{Lei85}
\begin{eqnarray}
&&\hspace{-0.9cm}H_{\rm c.m.}=\frac 1{2N_{s}m}\big({\bm P}-N_{s}e{\bm A}({\bm
R})\big)^2-N_{s}e({\bm E_0}+{\bm E}_t)\cdot {\bm R},\\
&&\hspace{-0.5cm}H_{\rm er}=\sum_{j}\left\{\frac{1}{2m}\Big[{\bm p}_{j\|}'-e{\bm A}
({\bm r}_{j}')\Big]^{2}+\frac{p_{jz}^2}{2m_z}+V(z_j)\right\}\nonumber\\
&&\hspace{1.1cm}+\sum_{i<j}V_{\rm c}({\bm r}_{i}'-{\bm r}_{j}',z_i,z_j),\,\,\,\,\, \label{her}
\end{eqnarray}
together with electron-impurity and electron-phonon couplings $H_{\rm ei}$ and $H_{\rm ep}$.
Here ${\bm A}({\bm r})=(-By,0)$ is the in-plane component of the vector potential of the perpendicular magnetic 
field,
${\bm E}_0$ and ${\bm E}_t$ are the dc and ac components of the uniform electric field inside,
$m$ and $m_z$ are, respectively, the electron effective mass parallel and perpendicular 
to the plane, $V(z)$ and $V_c({\bm r}_{i}'-{\bm r}_{j}',z_i,z_j)$ stand for 
the confined and Coulomb potentials.

The separation of the electron Hamiltonian into a c.m. part 
and a relative electron part amounts to look at electrons 
in the reference frame moving with their center of mass.   
The most important feature of this separation is that a spatially uniform electric field 
shows up only in $H_{\rm c.m.}$, and that $H_{\rm er}$ is the Hamiltonian of a many particle system 
subject to a magnetic field {\it without the electric field}.
This enables to deal with the relative electrons in the magnetic field
{\it without tilting the Landau levels}. 
 
We proceed with
the Heisenberg operator equations for the rate of change in the c.m. velocity 
${\bm V}=-i[{\bm R},H]$: $\dot{\bm V}=-i[{\bm V},H]$, and that of the relative electron energy $H_{\rm er}$:
$\dot{H}_{\rm er}=-i[H_{\rm er},H]$.
The c.m. coordinate ${\bm R}$ 
and velocity ${\bm V}$ in these equations can be treated classically, i.e., 
as the time-dependent
expectation values of the c.m. coordinate and velocity,
${\bm R}(t)$ and ${\bm V}(t)$, such that ${\bm R}(t)-{\bm R}(t^{\prime})
=\int_{t^{\prime}}^t{\bm V}(s)ds$.
In the case of the steady transport
under a modest radiation of single frequency $\omega$
 it suffices to assume that the c.m. 
velocity, i.e., the electron drift velocity, consists of a dc
part ${\bm v}$ and a stationary time-dependent part of the form
\begin{equation}
{\bm V}(t)={\bm v}+{\bm v}_1 \cos(\omega t)+{\bm v}_2 \sin(\omega t).
\end{equation}

For high mobility and high carrier density 2D systems at low temperatures
where effects of electron-impurity and electron-phonon scatterings are weak in comparison with
internal thermalization of this many electron system, it is a good 
approximation to carry out the statistical average of the above operator equations  
to leading orders in $H_{\rm ei}$ and $H_{\rm ep}$.
For this purpose we only need to know the distribution of relative electrons 
without being perturbed by $H_{\rm ei}$ or $H_{\rm ep}$.
The distribution function of the system described by Hamiltonian (\ref{her}) without electric field, 
is an isotropic Fermi-type with a single temperature $T_{e}$. 
Such a statistical average of the above operator equations yields 
the following force and energy balance equations in the steady state:
\begin{equation}
N_{s}e{\bm E}_{0}+N_{s} e ({\bm v} \times {\bm B})+
{\bm F}=0,\label{eqforce}
\end{equation}
\begin{equation}
N_{s}e{\bm E}_0\cdot {\bm v}+S_{p}- W=0.
\label{eqenergy}
\end{equation}
In this,
\begin{equation}
{\bm F}=\sum_{{\bm q}_\|}\left| U({\bm q}_\|)\right| ^{2}
\sum_{n=-\infty }^{\infty }{\bm q}_\|{J}_{n}^{2}(\xi ){\it \Pi}_{2}
({\bm q}_\|,\omega_0-n\omega )\label{exf0}
\end{equation}
is the time-averaged damping force against CM motion, $S_{p}$ is the time-averaged rate of the 
electron energy-gain from the ac field, having an expression obtained from Eq.\,(6) by replacing
${\bm q}_{\|}$ factor by $n\omega$. 
In Eq.\,(\ref{exf0}), $U({\bm q}_\|)$ is the effective impurity potential, $J_{n}(\xi )$ 
is the Bessel function of order $n$, ${\it \Pi}_2({\bm q}_\|,{\it \Omega})$ 
is the imaginary part of the electron density correlation function 
at electron temperature $T_{e}$ in the presence of the magnetic field, 
and $\omega_0\equiv{\bm q}_\|\cdot {\bm v}$. The argument in $J_n(\xi)$
is $\xi\equiv \sqrt{({\bm q}_\|\cdot {\bm v}_1)^2+({\bm q}_\|\cdot {\bm v}_2)^2}/\omega$.
Note that, although contributions of phonon scattering to ${\bm F}$ and $S_{p}$
are neglected here, it provides the main channel for electron energy dissipation to the lattice 
with a time-averaged energy-loss rate $W$, having the expression as given in Ref.\,\onlinecite{Lei03}. 

The ac components ${\bm v}_1$ and ${\bm v}_2$ of the electron drift velocity should be determined 
selfconsistently from the incident ac field ${\bm E}_{\rm is}$ by the Maxwell equations 
connecting both sides of the 2D system, taking account the scattering-related  
damping forces ${\bm F}_s$ and ${\bm F}_c$. \cite{Lei03} However, for the high-mobility systems 
at low temperatures, the effects of these scattering-related forces are much weaker than 
that of radiative decay \cite{Mikhailov04} and always negligible,
hence ${\bm v}_1$ and ${\bm v}_2$ are directly determined by the setup of the 2D system 
in the sample substrate.\cite{Lei03}

The effect of interparticle Coulomb interaction is included in the density correlation function 
to the degree of electron level broadening and screening 
(considered in the effective impurity and phonon potentials).
The remaining ${\it \Pi}_2({\bm q}_{\|}, {\it \Omega})$ function in Eqs.\,(\ref{exf0})
is that of a noninteracting  2D electron gas  
in the magnetic field, which can be written in the Landau representation as \cite{Ting}
\begin{eqnarray}
&&\hspace{-0.7cm}{\it \Pi}_2({\bm q}_{\|},{\it \Omega}) =  \frac 1{2\pi
l_{B}^2}\sum_{n,n'}C_{n,n'}(l_{B}^2q_{\|}^2/2) 
{\it \Pi}_2(n,n',{\it \Omega}),
\label{pi_2q}\\
&&\hspace{-0.7cm}{\it \Pi}_2(n,n',{\it \Omega})=-\frac2\pi \int d\varepsilon
\left [ f(\varepsilon )- f(\varepsilon +{\it \Omega})\right ]\nonumber\\
&&\,\hspace{2cm}\times\,\,{\rm Im}G_n(\varepsilon +{\it \Omega})\,{\rm Im}G_{n'}(\varepsilon ),
\label{pi_2ll}
\end{eqnarray}
where $l_{B}=\sqrt{1/|eB|}$ is the magnetic length,
$
C_{n,n+l}(Y)\equiv n![(n+l)!]^{-1}Y^l{\rm e}^{-Y}[L_n^l(Y)]^2
$
with $L_n^l(Y)$ the associate Laguerre polynomial, $f(\varepsilon
)=\{\exp [(\varepsilon -\mu)/T_{e}]+1\}^{-1}$ is the Fermi 
function at electron temperature $T_{e}$, 
and ${\rm Im}G_n(\varepsilon )$ is the density-of-states (DOS) function of the broadened Landau level $n$.

The LL broadening results from impurity, phonon and electron-electron scatterings.
In the experimental ultraclean GaAs-based 2D systems having mobility higher than $10^{3}$\,m$^{2}$/V\,s
the dominant elastic scatterings come mainly from residual impurities or defects in the background
rather than from remote donors,\cite{Umansky} and phonon and electron-electron scatterings are 
generally also not long-ranged because of the screening. On the other hand, since
the microwave-induced magnetoresistance oscillations occur at low temperatures and low magnetic fields
in high electron density samples, the cyclotron radius of electrons involving
in transport is generally much larger than the correlation length or the range 
of the dominant scattering potentials. In this case, the broadening of the LLs is expected  
to be a Gaussian form [$\varepsilon_n=(n+\frac{1}{2})\omega_c$ is the center of the $n$th Landau 
level],\cite{Raikh-1993}
\begin{equation}
{\rm Im}G_n(\varepsilon)=-(2\pi)^{\frac{1}{2}}{\it \Gamma}^{-1}
\exp[-2(\varepsilon-\varepsilon_n)^2/{\it \Gamma}^2]
\label{gauss}
\end{equation}
with a $B^{1/2}$-dependent half width expressed as 
\begin{equation}
{\it \Gamma}=(2\omega_c/\pi \tau_s)^{1/2}, 
\label{gamma12}
\end{equation} 
in which $\tau_s$, the single-particle lifetime or quantum scattering time in the zero magnetic field, depends on  
impurity, phonon and electron-electron scatterings. 

Expressions (\ref{gauss}) and (\ref{gamma12}) for the DOS of the $n$th level will be used 
in both the separated and overlapped LL regimes.
The total DOS of a 2D system of unit area in the magnetic field,  $g(\varepsilon)=-\sum_n {\rm 
Im}G_n(\varepsilon)/\pi^2 l_B^2$,
can be written as the sum of the DOS without magnetic field, $g_0$,
and a magnetic-field induced oscillatory part.
When the LL widths $2{\it \Gamma}$ are much larger than the level spacing $\omega_c$ 
(heavily overlapped LLs), the oscillatory part is much smaller than $g_0$
and one can keep only its fundamental harmonic component that
\begin{equation}
g(\varepsilon)\approx g_0\left[1-2\delta\cos(2\pi\varepsilon/\omega_c)\right],
\end{equation}
with $\delta=\exp(-\pi^2{\it \Gamma}^2/2\omega_c^2)=\exp(-\pi/\omega_c\tau_s)$.\cite{Coleridge-97} 

A reliable evaluation of single-particle lifetime $\tau_s$ (or the LL width ${\it \Gamma}$), 
which must include effects of impurity, phonon and electron-electron scatterings,
has not yet been available.  We treat it as an empirical parameter, or,
relate it to the zero-field linear mobility $\mu_0$ or the transport relaxation time $\tau_{\rm tr}$
using an empirical parameter $\alpha$ by  $1/\tau_s=4\alpha/\tau_{\rm tr}=4 \alpha e/m \mu_0$.

\section{Nonlinear resistivity and multiphoton processes}

These formulations are convenient for current-driven nonlinear magnetotransport 
of any configuration.
For an isotropic system where the frictional force ${\bm F}$ is in the opposite direction of 
the drift velocity ${\bm v}$
and the magnitudes of both the frictional force and the energy-loss rate depend only on 
$v\equiv |{\bm v}|$, we can write ${\bm F}({\bm v})=F(v){\bm v}/v$ and 
$W({\bm v})=W(v)$.  
In the Hall configuration with velocity ${\bm v}$ in the $x$ direction
${\bm v}=(v,0,0)$ or the current density $J_x=J=N_{s}ev$ and $J_y=0$,
Eq.\,(\ref{eqforce}) yields 
the transverse resistivity $R_{yx}=B/N_{s}e$, and the longitudinal resistivity $R_{xx}$ and 
differential resistivity $r_{xx}$ at given $v$ as
\begin{eqnarray} 
&&R_{xx}= -F(v)/(N_{s}^2e^2v), \label{eqrxx}\\
&&r_{xx}=-({\partial F(v)}/{\partial v})/(N_{s}^2e^2). \label{eqdr}
\end{eqnarray}

In the linear case, the resistivity $r_{xx}$ oscillations (peak-valley
pairs) result from the resonant real photon-assisted electron transitions between different 
Landau levels. \cite{Lei06}
We denote a real-photon assisted process in which an electron jumps across $l$ 
LL spacings
with the assistance (emission or absorption) of $n$ photons as $n\omega$:$l\omega_c$, 
or $n$:$l$. This process contributes, 
in the $r_{xx}$--$\omega/\omega_c$ curve,
a pair structure consisting of a maximum and a minimum on both sides of $\omega/\omega_c=l/n$.

Thus, the single-photon process $1$:$1$, the two-photon process
$2$:$2$, the three-photon process $3$:$3$,... 
all contribute to the maximum-minimum pair around $\omega/\omega_c=1$;
the two-photon process $2$:$1$, the four-photon process
$4$:$2$,... 
all contribute to the maximum-minimum pair around $\omega/\omega_c=1/2$;
the three-photon process $3$:$1$, the six-photon process
$6$:$2$,... 
all contribute to the maximum-minimum pair around $\omega/\omega_c=1/3$;  
etc. The virtual photon process (zero-photon process) during which the electron may 
absorb (emit) one or several photons and then emits (absorbs) them before 
finishing an intra-LL transition (assuming separated LLs for convenience), 
which can be denoted as $0$:$0$, gives rise to an overall resistivity suppression
without structure. \cite{Lei06}

The effects of a finite current show up as the energy (frequency) shift 
$\omega_0={\bm q}_{\|}\cdot {\bm v}$
in the density-correlation function  ${\it \Pi}_2({\bm q}_{\|},{\it \Omega})$ in Eq.\,(\ref{exf0}).
This energy shift indicates that an electron, having momentum ${\bm q}_{\|}$ and participating in 
the system integrative drift motion of velocity ${\bm v}$, carries an extra energy 
${\bm q}_{\|}\cdot {\bm v}$ during its transition.
 In the case of low temperature 
($T_{e}$ much less than the Fermi energy $\varepsilon_{F}$)  
and large Landau-level filling factor ($\nu=\varepsilon_{F}/\omega_c\gg 1$),
the major contributions to the summation in Eq.\,(\ref{pi_2q})
come from terms $n\sim n'\sim \nu$, then the function $C_{n,n'}(x)$ has a sharp principal maximum
near $x=4\nu$. Therefore, as a function of the in-plane momentum $q_{\|}$,
 ${\it \Pi}_2({\bm q}_{\|},{\it \Omega})$ sharply peaks around $q_{\|} \simeq 2K_{F}$. 
As the composite effect of this energy shift, 
the electron possesses an extra energy 
\begin{equation}
\omega_j\equiv 2K_{F}v=\sqrt{8\pi/N_{s}}J/e
\end{equation}
for its transition in addition to those provided by photons. 

Therefore, the resonant condition applied in the linear case 
for electron scattering across $l$ LL spacings with the assistance (emission or absorption) 
of $n$ photons, $n\omega= l\omega_c$, should change to 
\begin{equation}
\omega_j\pm n\omega=\pm l\omega_c\,\,\,(n=0,1,2,...,\,\,  l=0,1,2,...) \label{reson}
\end{equation}
when the system has
an integrative motion of velocity $v$, i.e., a finite current density $J=N_{s}e v$.
We can still use symbol $n$:$l$ to denote this process. We will also use symbol
$n$- to represent all the $n$-photon participating processes (electron may jump across any LL spacings).

Increase in the current density from zero reduces the rates of the scattering processes
resonant in the linear case, such as those  
of zero-photon processes $0$:$0$ everywhere, 
of single-photon process $1$:$1$ and two-photon process $2$:$2$ at $\omega/\omega_c=1$, 
of two-photon processes $2$:$1$ and four-photon processes $4$:$2$ at $\omega/\omega_c=1/2$. 
But it may increase the rates of the scattering processes originally absent or very small 
in the linear case, such as those of single-photon process $1$:$1$ and three-photon process $3$:$1$ at 
$\omega/\omega_c=1/2$.

At cyclotron resonance  and its harmonics $\omega/\omega_c=1,2,4,...$, the resonant condition
(\ref{reson}) can be satisfied only at 
\begin{equation}
\epsilon_j\equiv\omega_j/\omega_c=0, 1, 2,...,
\end{equation} 
where the scattering rate peaks and the differential resistivity exhibits maximum. 
Therefore, at these magnetic fields the $r_{xx}$ oscillation with changing $\epsilon_j$ always has 
a periodicity $\Delta\epsilon_j\approx 1$.

\begin{figure}
\includegraphics [width=0.45\textwidth,clip=on] {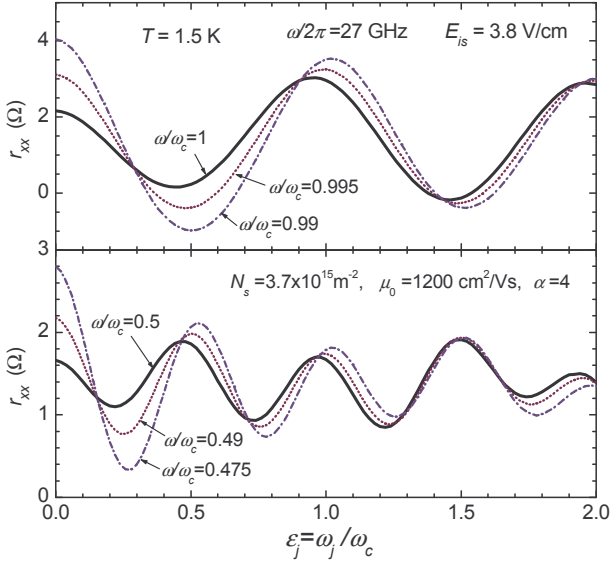}
\vspace*{-0.2cm}
\caption{(Color online) Magnetoresistivity $r_{xx}$ vs $\epsilon_{j}=\omega_j/\omega_c$
at $\omega/\omega_c=1,\,0.995,\,0.99,\,0.5,\,0.49$ and 0.475.}
\label{fig1}
\end{figure}
\begin{figure}
\includegraphics [width=0.45\textwidth,clip=on] {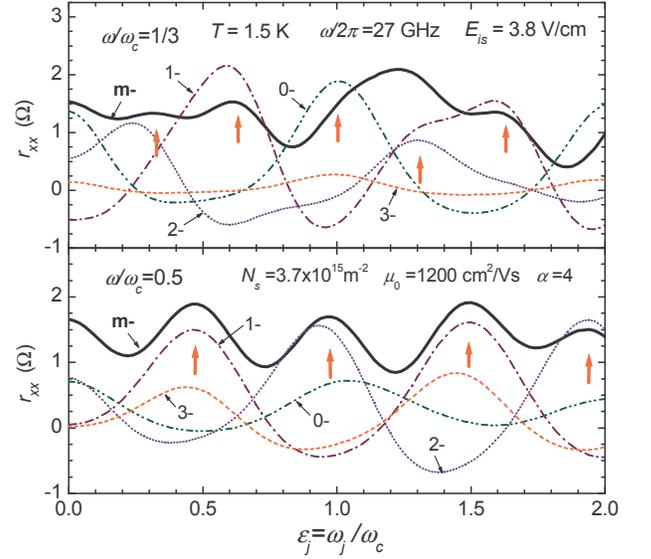}
\vspace*{-0.2cm}
\caption{(Color online) Magnetoresistivity $r_{xx}$ vs $\epsilon_j$
at $\omega/\omega_c=1/3$ and 0.5. Curves with 0-, 1-, 2-, 3- and m- are separated contributions from
zero-photon, single-photon, two-photon, three-photon and  multiphoton processes.}
\label{fig2}
\end{figure}

The situation changes near subharmonic cyclotron resonant fields.
For instance, at $\omega/\omega_c=1/2$, the resonant condition (\ref{reson}) 
for zero-photon, two-photon, and four-photon processes ($n=0, 2$ and 4) is satisfied at $\epsilon_j=0,1,2,...$,
indicating these even-photon processes contribute resistivity maxima there. 
At $\epsilon_j=0.5,1.5,...$, on the other hand, the resonant condition (\ref{reson}) is satisfied 
for single-photon and three-photon processes ($n=1$ and 3), indicating these odd-photon processes contribute 
resistivity maxima. Because of this, the period
of $r_{xx}$ oscillation with $\epsilon_j$ may shrink to $\Delta\epsilon_j\approx 0.5$, or  
the oscillation frequency doubles, at $\omega/\omega_c=1/2$.

At the third subharmonic field, $\omega/\omega_c=1/3$, in addition to the appearance of resonant scattering 
of zero-photon and three-photon processes at $\epsilon_j=0,1$ and 2, the resonance scattering can also 
appear at $\epsilon_j=1/3, 4/3$ and 5/3 for two-photon processes and at $\epsilon_j=2/3, 4/3$ and 5/3 
for single-photon processes, leading to possible shrinkage of the $r_{xx}$ oscillation period 
to $\Delta\epsilon_j\approx 1/3$, or the frequency tripling. 

These analyses are confirmed by the numerical results shown in Figs.\,1 and 2, where we plot
the differential resistivity $r_{xx}$ 
calculated from Eq.\,(\ref{eqdr})
as a function 
of the normalized current density $\epsilon_j=\omega_j/\omega_c$
for a GaAs-based system of $N_{s}=3.7\times 10^{15}$\,m$^{-2}$ and 
$\mu_0=1200$\,m$^2$/V\,s at $T=1.5$\,K, irradiated by a
 $27$\,GHz microwave having incident electric field amplitude $E_{\rm is}=3.8$\,V/cm.
The elastic scatterings are due to a mixture
of short-range and background impurities, and the broadening 
 parameter $\alpha=4$, or the LL width $2{\it \Gamma}/\omega_c\simeq 0.72(\omega/\omega_c)^{1/2}$,
 in the separated LL regime for $\omega/\omega_c<1$. The predicted $r_{xx}$ behaviors
  near $\omega/\omega_c=1$ and $\omega/\omega_c=0.5$
 are in excellent agreement with recent experimental findings.\cite{Hatke08-2}
 The $r_{xx}$ oscillations at $\omega/\omega_c=1/3$ exhibit the observable frequency tripling
 under this strength of microwave irradiation.

\section{Summary}

In summary, the electron in a dc-driven system having an integrative drift velocity $v$ carries  
an extra average energy $\omega_j=2K_{F}v$ in its transition between LLs, 
giving rise to a current-dependent resonance condition. 
The peaks of the differential resistivity $r_{xx}$ arise from resonant
multiphoton assisted scatterings of electron jumping across $l$  
LL spacings. 
The current-induced consecutive emergence of different resonant multi-photon processes
results in the frequency doubling and tripling of the resistance oscillations 
near the 2nd and 3rd subharmonic fields.

\vspace{0.1cm}

The author is thankful to M. Zudov for helpful discussions.
This work was supported by the projects of the National Science Foundation of China,
the Special Funds for Major State Basic Research Project (No. 2007CB310402), 
and the Shanghai Municipal Commission 
of Science and Technology (No. 06dj14008).

\end{document}